\documentclass[10pt]{article}
\usepackage{amsmath,amsfonts,subfig,graphicx,lineno,indentfirst, textcomp}
\usepackage{authblk}
\usepackage[margin=2.5cm]{geometry}
\title{Deep learning approach for flow visualization in background oriented schlieren}

\author[1]{Viren S Ram}
\author[2]{Tullio de Rubeis}
\author[3]{Dario Ambrosini}
\author[1,4,*]{Rajshekhar Gannavarpu}

\affil[1]{Department of Electrical Engineering, Indian Institute of Technology Kanpur, Kanpur-208016, India }
\affil[2]{
    DICEAA, University of L'Aquila, P. le E. Pontieri 1, L'Aquila, 67100, Italy}
\affil[3]{DIIIE, University of L'Aquila, P. le E. Pontieri 1, L'Aquila, 67100, Italy}
\affil[4]{Center for Lasers and Photonics, Indian Institute of Technology Kanpur, Kanpur-208016, India}
\affil[*]{Corresponding author: gshekhar@iitk.ac.in}

\begin{document}
\maketitle
\begin{abstract}
Diffractive optical element based background oriented schlieren (BOS) is a popular technique for quantitative flow visualization.
This technique relies on encoding spatial density variations of the test medium in the form of an optical fringe pattern; and hence, its accuracy is directly influenced by the quality of fringe pattern demodulation. 
We introduce a robust deep learning assisted subspace method which enables reliable fringe pattern demodulation even in the presence of severe noise and uneven fringe distortions in recorded BOS fringe patterns.
The method's effectiveness to handle fringe pattern artifacts is demonstrated via rigorous numerical simulations. 
Furthermore, the method's practical applicability is experimentally validated using real-world BOS images obtained from a liquid diffusion process.
\end{abstract}

\section{Introduction}
Flow analysis of diffusion phenomena \cite{cussler2009diffusion, tyrrell2013diffusion} has vital applications in different domains such as  biology \cite{gilpin2017flowtrace,yu2005diffusion}, fuel studies \cite{yeganeh2023visualization,hu2024visualization}, chemical engineering \cite{itsariyapinyo2022visualization,wijmans1995solution}, and environmental sciences \cite{choy2017diffusion,wu2025accurate}. 
For diffusion measurements, optical interferometric methods offer several attractive features such as whole field measurement, non-invasive operation, good resolution and versatility for both qualitative and quantitative analysis \cite{ambrosini2008overview, prenel2012flow}. 
Notable optical techniques in this field include holographic interferometry \cite{ruiz1985holographic,chikode2021determination,anand2006diffusivity,he2015development}, electronic speckle pattern interferometry \cite{hassani2021applications,paoletti1997temperature, riquelme2007interferometric}, speckle decorrelation \cite{ambrosini2002speckle,jeon2019optofluidic}, shearing interferometry \cite{rashidnia2002development,xu2024dynamic} and phase-shifting interferometry \cite{torres2012development,maruyama2006situ,luo2020optofluidic}.
Over the years, diffractive optical element based background oriented schlieren \cite{schirripa2004liquid, settles2017review,ramaiah2020fast} has emerged as a popular optical technique for flow metrology.
The primary advantages of the BOS technique include simple experimental configuration and robust design \cite{raffel2015background}.
The core principle behind most of these techniques is reliable fringe pattern demodulation or accurate extraction of the fringe pattern's phase map, since it directly corresponds to refractive index variations induced by the diffusion process. 
Nevertheless, phase retrieval is inherently challenging due to factors such as noise, uneven fringe intensity fluctuations and the requirement of number of frames for fringe processing, and accordingly, various approaches have been explored in the literature. 
Phase-shifting \cite{servin2009general,vinnichenko2025background} is a widely used phase measurement technique; however,  its reliance on recording and processing multiple phase-stepped images presents experimental challenges for studying dynamic processes.
In contrast, fringe analysis techniques such as Fourier transform \cite{takeda1982fourier}, windowed Fourier transform \cite{kemao2007two,kemao2004windowed}, and wavelet transform \cite{watkins1999determination,watkins2012review} operate on single fringe pattern for phase extraction.
Among these, the Fourier transform method is widely popular for flow visualization \cite{spagnolo1994fourier, vinnichenko2023performance,blanco2016performance} primarily due to its operational ease and computational simplicity. 
However, since the Fourier transform is a global operation, its effectiveness is significantly impacted by localized fringe distortions and noise within the fringe pattern.

The primary objective of proposed work is to develop a fringe pattern demodulation method for flow visualization which addresses the dual problems of noise corruption and non uniform fringe intensity fluctuations. 
To achieve this, we introduce deep learning assisted signal subspace approach for robust phase retrieval under extreme noise conditions coupled with non-linear modulation variations.
Of late, deep learning approaches have gained significant attention in optical fringe analysis, enabling a wide range of applications \cite{feng2019fringe, wang2019net, zhang2018fast, park2023fast, vithin2022phase, yan2019fringe, gurrola2022u, zeng2021deep, narayan2023deep, pandey2023non, mohammed2018net, ram2024fringe}. 
The ability to learn complicated patterns and representations from the training datasets makes deep learning a powerful approach for processing complex fringe patterns; and our work aligns with this trend to enable significant advancements in BOS based flow visualization.

The paper is organized as follows.
 Section \ref{sec:theory} explains the working of the proposed method. 
Subsequently, the utility of proposed method for processing numerically simulated fringe patterns is shown via different validation metrics in Section \ref{sec:sim}. 
Further, in Section \ref{sec:exp}, the practical applicability of the proposed method is tested for processing experimental images obtained for diffusion experiment in BOS. 
In the end, concluding remarks are presented.

\section{Theory}

\label{sec:theory}
The general mathematical form of a fringe pattern recorded in diffractive optical element based BOS can be expressed as \cite{schirripa2004liquid}
\begin{equation}
    i_r(x,y) = i_0(x,y) + i_1(x,y)\cos(2\pi f_x x + \phi(x,y))
\end{equation}
where $(x,y)$ denote the spatial coordinates, $i_r$ is the recorded intensity, $i_0$ denotes the background intensity, $i_1$ denotes the modulation or fringe amplitude term and $f_x$ is the spatial carrier frequency introduced by the grating projection.
In fringe analysis, the primary aim is to estimate the phase map $\phi$ from the above equation.
This becomes a challenging problem in the presence of non-uniform fluctuations in background term $i_0$ or modulation term $i_1$, which could be caused by irregular illumination, optical component inhomogeneities and imaging artifacts.
This problem is further constrained when noise is incorporated in the imaging model.

\begin{figure}[t!]
\centering
    \includegraphics[scale=0.4]{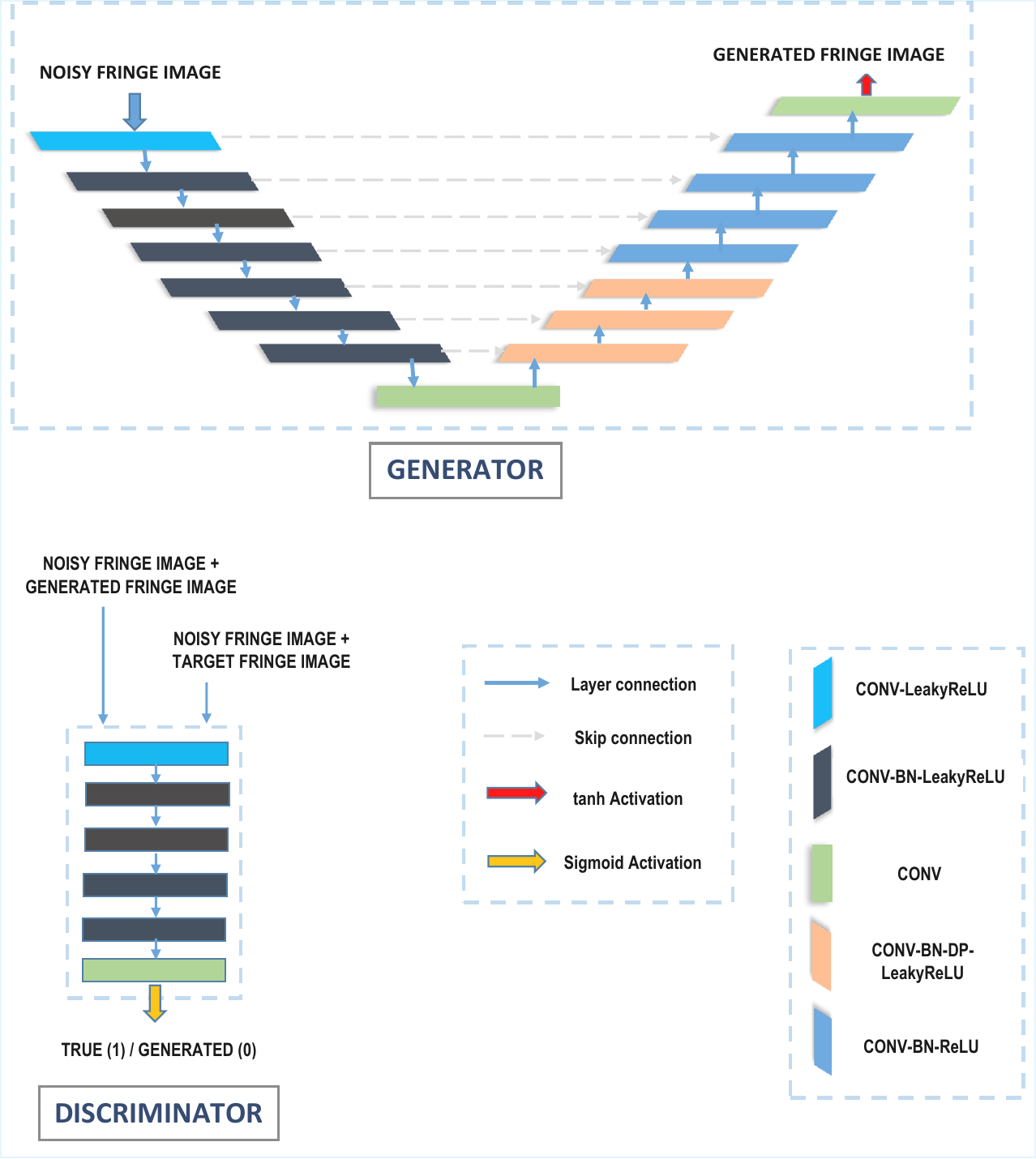}
    \caption{Schematic of deep learning model}
    \label{fig:ml_schematic}
\end{figure}

In this work, we propose deep learning assisted subspace processing method for phase estimation. 
The deep learning model plays a crucial role in preprocessing the fringe patterns using fringe normalization, which ensures that the effect of undesirable intensity distortions and artifacts is minimized, improving the overall signal quality. 
Further, the subspace approach leverages the underlying structure of the signal, effectively separating the phase component from noise component.
Hence, by combining deep learning with the signal subspace approach, our proposed framework enhances both the accuracy and robustness for phase estimation, making it well-suited for analyzing diffusion processes in BOS setup.
In our method, we use conditional generative adversarial networks \cite{wang2018high, isola2017image, goodfellow2020generative} as our deep learning model. 
The schematic of the machine learning model is shown in Figure \ref{fig:ml_schematic}.
The model is comprised of two key components: a generator and a discriminator. These components engage in an adversarial training process, where the generator is pushed to produce images that increasingly resemble the ground truth.
Through this iterative process, the generator continuously refines its weights, learning to produce increasingly realistic images that closely match the features of the ground truth. 

The generator's primary objective is to create high-quality normalized fringe images that replicate the characteristics of the original ground truth dataset.
This is achieved using U-Net architecture, known for its effectiveness in image-to-image translation tasks.
The architecture leverages skip connections, which links the encoder and decoder \cite{ronneberger2015u}, preserving essential spatial information and features throughout the translation process.
In parallel, the discriminator sub-model is trained through adversarial techniques, with the primary aim of accurately distinguishing between authentic samples from the original training dataset and images generated by the model.
This ensures that the discriminator continuously learns to improve its performance in classifying images as either real or synthetic.
The binary PatchGAN classifier \cite{li2016precomputed} employed within the discriminator focuses on local image patches instead of each pixel values, enhancing its capability to identify real and false images.
For our model, the objective function is represented as \cite{isola2017image}
\begin{equation}
\label{eq:GANs}
\begin{aligned}
    \textbf{O} = \frac{1}{N} \sum_{m=1}^N \log(1-\textbf{D}(I_m),\textbf{G}(I_m)) + \log\textbf{D}(I_m,I_{gm}) \\
    + \lambda||I_{gm} - \textbf{G}(I_m)||_1
\end{aligned}
\end{equation}
which is minimized with respect to parameters controlling generator ($G$) and maximized for parameters governing the discriminator ($D$).
The hyperparameter $\lambda$ is a scaling factor used for training stabilization, $N$ denotes the number of images used for training, $I_m$ represents the non-normalized input fringe pattern, and $I_{gm}$ is its corresponding ground truth normalized fringe pattern. 
The notation $||.||_1$ refers to the $L_1$ norm. 
The generator and discrimator model is summarized in Tables \ref{table:Gen} and \ref{table:Dis} respectively.  

\begin{table}[t!]
\centering
    \caption{Summary of Generator sub-model (DL: Downsampling Layer; UL: Upsampling Layer)}
\begin{tabular}{llll} 
\hline
\bf S.No & \bf Layer & \bf Output Shape & \bf Inputs \\ [0.9ex] 
 \hline
 (1) & Input Layer & (256,256,1) &  \\ 
 (2) & DL-1 & (None,128,128,64) &  (1) \\
 (3) & DL-2 & (None,64,64,128) & (2)\\
 (4) & DL-3 & (None,32,32,256) & (3) \\
 (5) & DL-4 & (None,16,16,512) & (4) \\
 (6) & DL-5 & (None,8,8,512) & (5)\\
 (7) & DL-6 & (None,4,4,512) & (6) \\
 (8) & DL-7 & (None,2,2,512) & (7) \\  
 (9) & Bridge Layer & (None,1,1,512) & (8)\\
 (10) & UL-1 & (None,2,2,1024) & (9,8) \\
 (11) & UL-2 & (None,4,4,1024) & (10,7)\\
 (12) & UL-3 & (None,8,8,1024) & (11,6) \\
 (13) & UL-4 & (None,16,16,1024) & (12,5)\\
 (14) & UL-5 & (None,32,32,1024) & (13,4)\\
 (15) & UL-6 & (None,64,64,1024) & (14,3)\\
 (16) & UL-7 & (None,128,128,128) & (15,2)\\
 (17) & Output Layer & (None,256,256,1) & (16)\\
 \hline
\end{tabular}
\label{table:Gen}
\end{table}

\begin{table}[h!]
\centering
    \caption{Summary of Discriminator sub-model (DL: Downsampling Layer)}
\begin{tabular}{llll} 
\hline
\bf S.No & \bf Layer & \bf Output Shape  & \bf Inputs \\ [0.9ex] 
 \hline
 (1) & Input Layer & (256,256,2) & \\ 
 (2) & DL-1 & (None,128,128,64) & (1) \\
 (3) & DL-2 & (None,64,64,128) & (2)\\
 (4) & DL-3 & (None,32,32,256)  & (3) \\
 (5) & DL-4 & (None,16,16,512) & (4) \\
 (6) & DL-5 & (None,16,16,512) & (5)\\
 (7) & DL-6 & (None,16,16,1) & (6) \\
 \hline
\end{tabular}
\label{table:Dis}
\end{table}

Next, we apply Hilbert transform \cite{ikeda2005hilbert} on the fringe pattern obtained from deep learning model to generate the analytic fringe signal of the form 
\begin{equation}
\label{eq:analytic}
    i_a(x,y) = e^{j\phi(x,y)}+\eta(x,y)
\end{equation} 
where $\phi$ is the phase map and $\eta$ is the noise term, assumed to be additive white Gaussian noise.
Next, we consider a small window region where the fringe signal is approximated as a linear phase signal.
Let $i_w$ and $\phi_w$ denote the signal and phase within  the window, which are given as
\begin{equation}
\label{eq:win_eq}
    i_w(x,y) = e^{j\phi_w(x,y)}+ \eta_w(x,y)
\end{equation} 
\begin{equation}
\label{eq:win_ph}
\phi_w(x,y) = a_0 + a_1x + a_2y
\end{equation} 
where $x,y \in [-W,W]$, $S=2W + 1$ is the window size parameter, $(a_1, a_2)$ represent first order polynomial coefficients and $a_0$ denotes zero order polynomial coefficient.
Consequently, the phase at each pixel can be calculated by estimating the unknown polynomial coefficients.
This is achieved by using the rotational invariance property of the signal subspace \cite{roy1989esprit,ramaiah2019graphics} of the given signal.
Note that Eq.(\ref{eq:win_eq}) can be rewritten in matrix form as
\begin{equation}
    \label{eq:win_matrix}
    i_w = \textbf{paq}^T + \textbf{N}_\eta
\end{equation} 
where $\textbf{N}_\eta$ is the noise matrix, $[.]^T$ denotes the transpose operation and,
\begin{equation}
\begin{aligned}
\textbf{p} &= \begin{bmatrix} e^{j(-W)a_1} \\ \vdots \\ e^{j(W-1)a_1} \\ e^{j(W)a_1}
\end{bmatrix},\;
\textbf{q} &= \begin{bmatrix} e^{j(-W)a_2} \\ \vdots \\ e^{j(W-1)a_2} \\ e^{j(W)a_2}
\end{bmatrix},\;
\textbf{a} &= \begin{bmatrix}  e^{ja_0}  \end{bmatrix}       
\end{aligned}
\end{equation}
Subsequently we perform singular value decomposition (SVD) operation \cite{golub1965calculating} which provides,
\begin{equation}
\label{eq:svd_eq}
i_w = \mathbf{U}\mathbf{\Sigma}\mathbf{V}^H = \sum_{m=1}^{S}\sigma_m\mathbf{u_m}\mathbf{v_m}^H 
\end{equation} 
where $[.]^H$ denotes complex conjugate transpose operation or Hermitian of matrix, $\textbf{u}_m$ denotes the left singular vectors or columns of $\textbf{U}$ whereas $\textbf{v}_m$ denotes the right singular vectors or columns of $\textbf{V}$.
Further, $\sigma_m $ indicates the singular values, which are arranged in decreasing order. 
The SVD provides an elegant approach to decompose the noisy data matrix $i_w$ in terms of a prominent signal component and remaining noise component.
In our analysis, the signal subspace corresponds to the first or dominant singular value and its associated left and right singular vectors.
Similarly, the rest of the singular values and singular vectors correspond to noise subspace.
In other words, we have
\begin{equation}
    \label{eq:svd_comp}
i_w(x,y) = \sigma_1\textbf{u}_1\textbf{v}_1^H + \sum_{m=2}^{M}\sigma_m\textbf{u}_m\textbf{v}_m^H
\end{equation} 
where the first term corresponds to signal and the second term depicts the noise component.
From Eq.(\ref{eq:win_matrix}) and Eq.(\ref{eq:svd_comp}), we note that the vectors $\textbf{p}$ and $\mathbf{u_1}$ span the same column space, and a similar observation is also true for vectors $\textbf{q}$ and $\mathbf{v_1}$.
Next, utilizing the signal subspace spanning vectors $\mathbf{u_1}$ and $\mathbf{v_1}$, we extract the polynomial coefficients $a_1$ and $a_2$ as given below \cite{roy1989esprit}
\begin{align}
a_2 &= arg(\mathbf{U_1^{+}U_2}) \\\notag
    a_1 &= arg(\mathbf{V_1^{+}V_2})
\end{align} 
\begin{equation}
\begin{aligned}
\mathbf{U_1} &= \begin{bmatrix} \mathbf{I}_{S-1} &  \mathbf{0}_{S-1}\end{bmatrix}\mathbf{u}_1\\
\mathbf{U_2} &= \begin{bmatrix} \mathbf{0}_{S-1} &  \mathbf{I}_{S-1}\end{bmatrix}\mathbf{u}_1\\
\mathbf{V_1} &= \begin{bmatrix} \mathbf{I}_{S-1} &  \mathbf{0}_{S-1}\end{bmatrix}\mathbf{v}_1^*\\
\mathbf{V_2} &= \begin{bmatrix} \mathbf{0}_{S-1} &  \mathbf{I}_{S-1}\end{bmatrix}\mathbf{v}_2^*
\end{aligned}
\end{equation} 
where $[.]^{+}$ is Moore-Penrose pseudo-inverse operation, $\mathbf{I}_{S-1}$ indicates Identity matrix of size $(S-1)$ and $\mathbf{0}_{S-1}$ is zero filled column matrix of size $(S-1)$, $[.]^*$ indicates the complex conjugation operation and $arg$ indicates the argument of a complex number obtained using arc-tangent operation.
Subsequently, the parameter $a_0$ can be estimated using,
\begin{equation}
a_0 = arg[\overline{e^{-j(a_1x + a_2y)}}]
\end{equation}
where $\overline{[.]} $ denotes the averaging operation. 
Once the parameters $a_0, a_1$ and $a_2$ are estimated, the phase can be calculated by substituting their values in equation (\ref{eq:win_ph}). 
The overall phase map for all pixels can be estimated in similar fashion by sliding the window over the entire fringe pattern. 
Due to the argument operator, we applied a phase unwrapping procedure \cite{herraez2002fast} to obtain the smooth phase map. 

\begin{figure}[t!]
	\centering
	\includegraphics[width=0.4\textwidth]{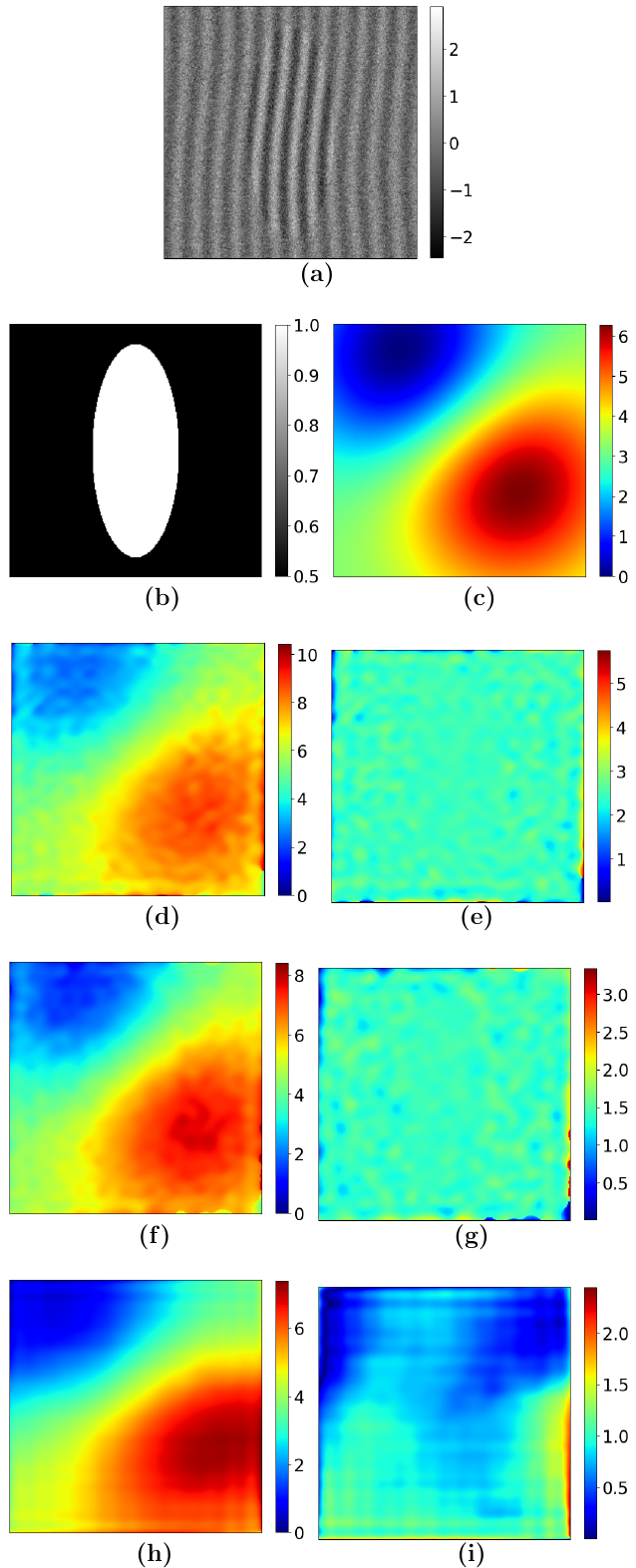} 
	\caption{
    (a) Noisy fringe pattern.
        (b) Type $M_1$ fringe modulation map.
        (c) Original phase map.
    (d) Phase estimated using Fourier transform method and (e) corresponding error.
    (f) Phase estimated using WFT method and (g) corresponding error.
    (h) Phase estimated using proposed method and (i) corresponding error.
    }
	\label{fig:para}
\end{figure}

\begin{figure}[t!]
	\centering
	\includegraphics[width=0.4\textwidth]{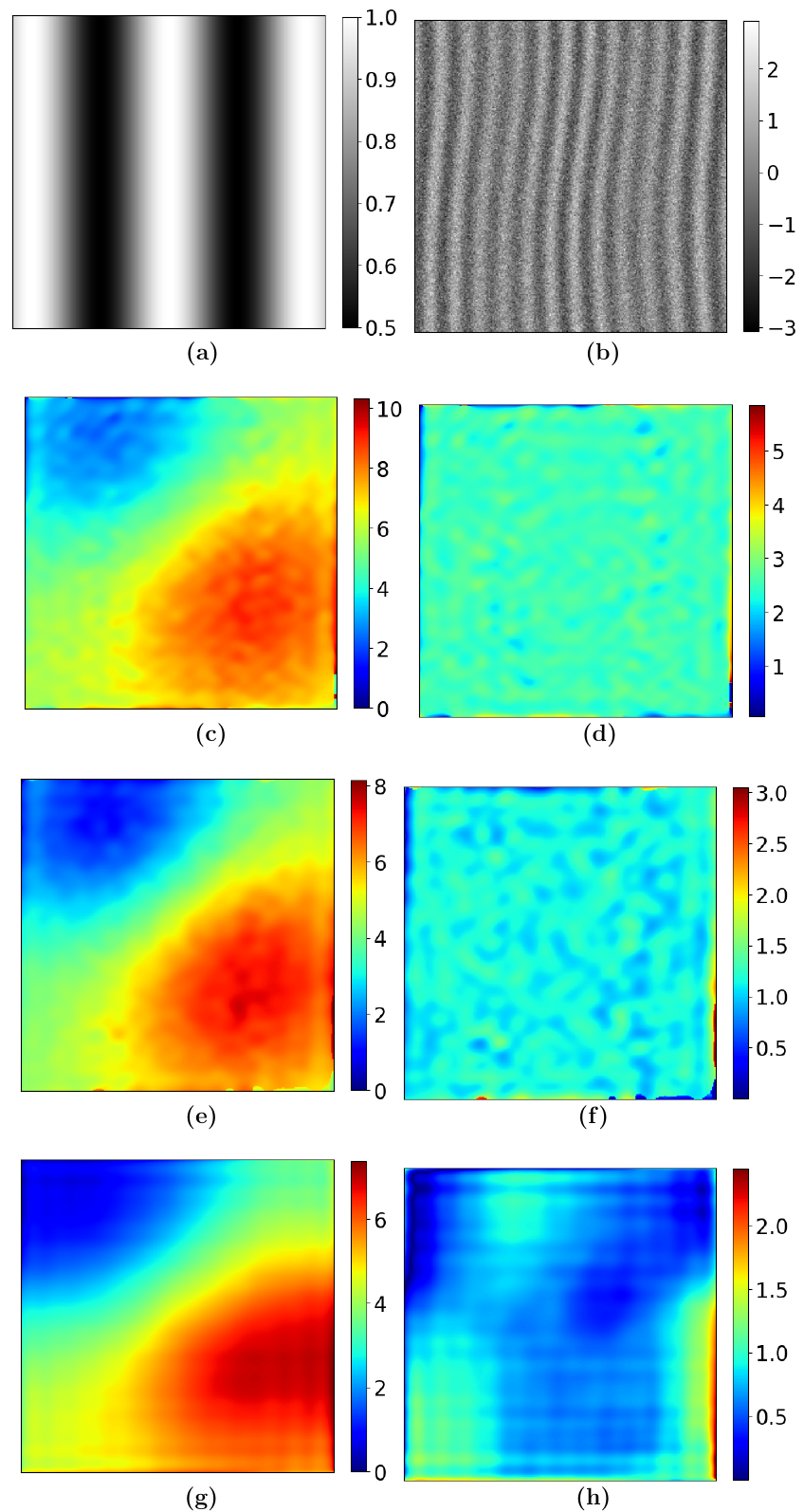} 
	\caption{
        (a) Type $M_2$ fringe modulation map.
    (b) Noisy fringe pattern.
    (c) Phase estimated using Fourier transform method.
    (d) Corresponding phase estimation error.
    (e) Phase estimated using WFT method.
    (f) Corresponding phase estimation error.
    (g) Phase estimated using proposed method.
    (h) Corresponding phase estimation error.
    }
	\label{fig:sinu}
\end{figure}

\begin{figure}[t!]
	\centering
	\includegraphics[width=0.45\textwidth]{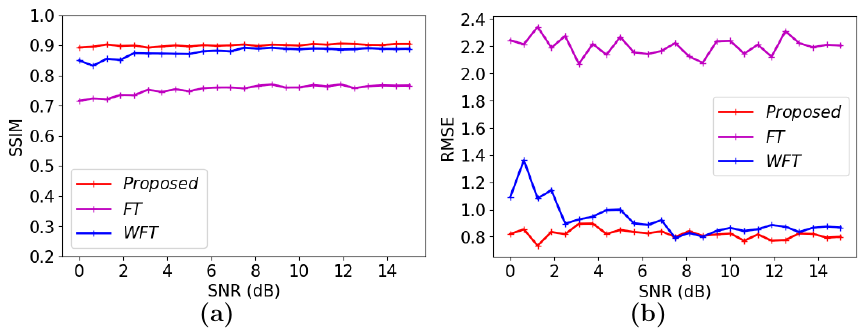} 
	\caption{(a) SSIM versus SNR and  (b) RMSE versus SNR for simulated fringe patterns with type $M_1$ fringe modulation }
	\label{fig:para_plot}
\end{figure}

\begin{figure}[h!]
	\centering
	\includegraphics[width=0.45\textwidth]{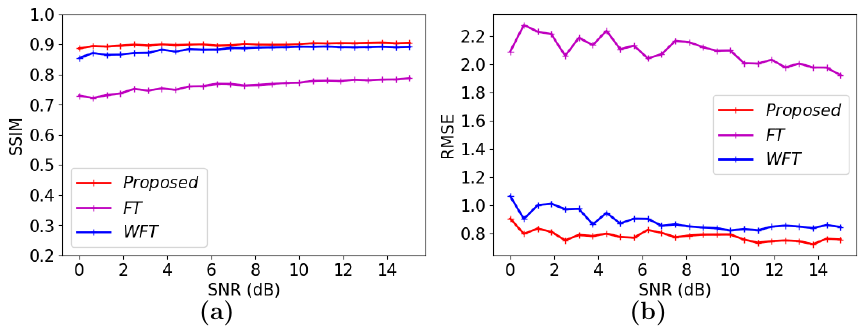} 
	\caption{(a) SSIM versus SNR (b) RMSE versus SNR for simulated fringe patterns with type $M_2$ fringe modulation}
	\label{fig:sinu_plot}
\end{figure}

\begin{figure}[t!]
	\centering
	\includegraphics[width=0.4\textwidth]{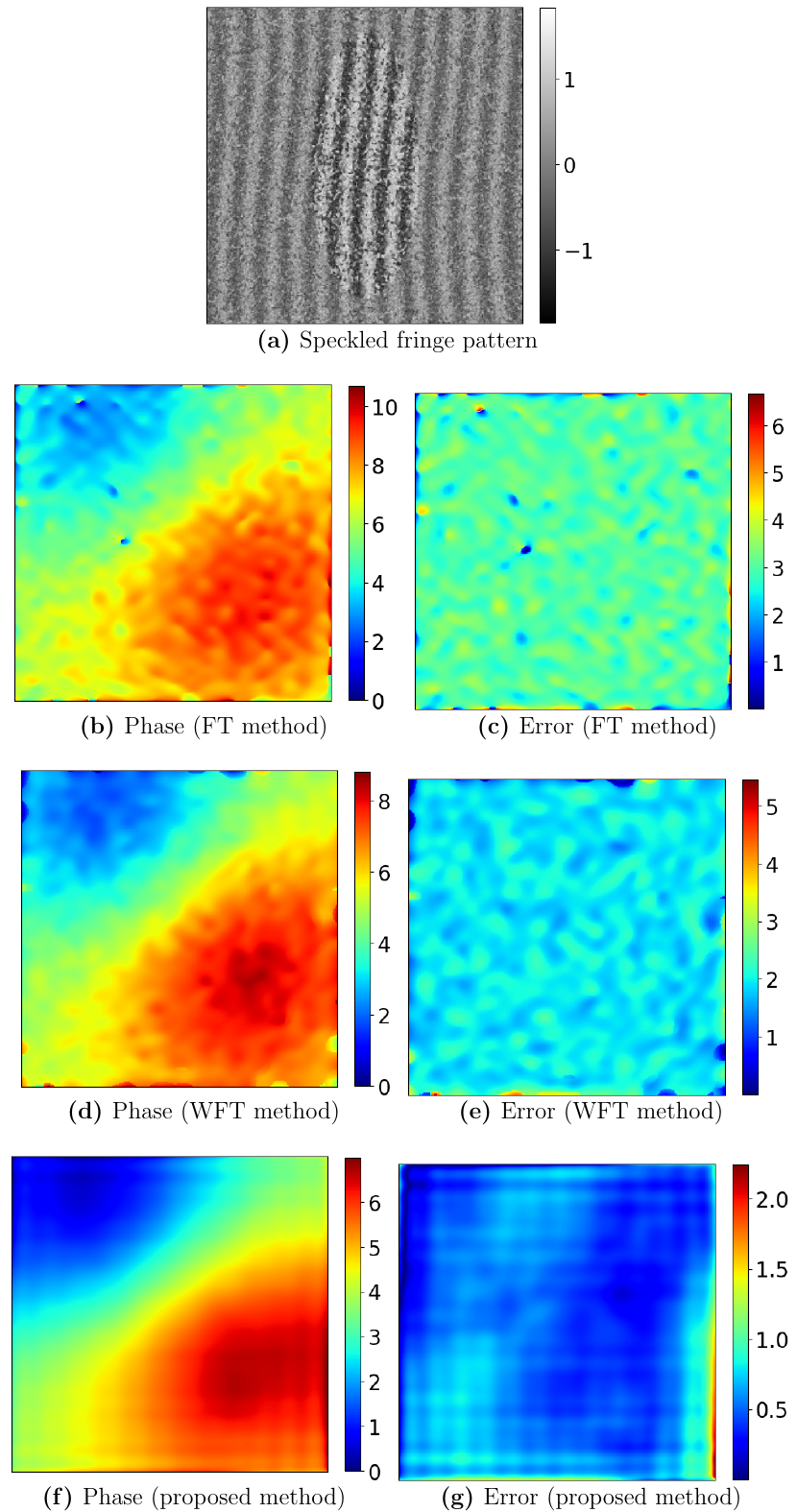} 
	\caption{
        (a) Speckle fringe pattern with type $M_1$ fringe modulation.
    (b) Phase estimated using Fourier transform method.
    (c) Corresponding phase estimation error.
    (d) Phase estimated using WFT method.
    (e) Corresponding phase estimation error.
    (f) Phase estimated using proposed method.
    (g) Corresponding phase estimation error.
}
	\label{fig:para_speckle}
\end{figure}

\begin{figure}[t!]
	\centering
	\includegraphics[width=0.4\textwidth]{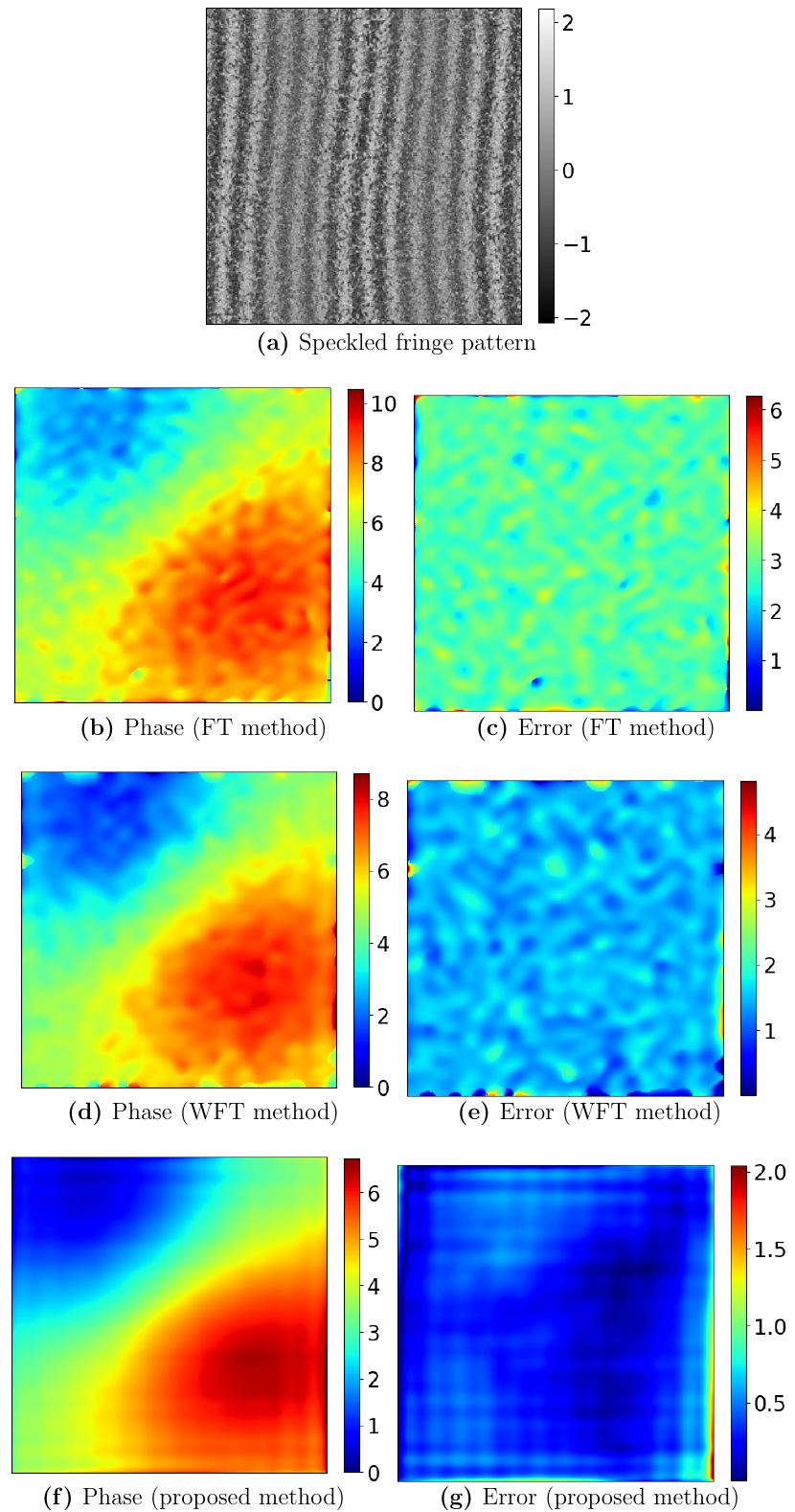} 
	\caption{
        (a) Speckle fringe pattern with type $M_2$ fringe modulation.
    (b) Phase estimated using Fourier transform method.
    (c) Corresponding phase estimation error.
    (d) Phase estimated using WFT method.
    (e) Corresponding phase estimation error.
    (f) Phase estimated using proposed method.
    (g) Corresponding phase estimation error.
}
	\label{fig:sinu_speckle}
\end{figure}
\begin{figure}[t!]
	\centering
	\includegraphics[width=0.45\textwidth]{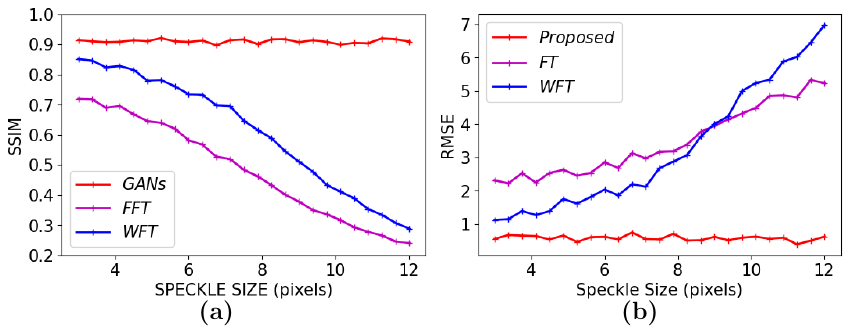} 
	\caption{(a) SSIM versus speckle size and (b) RMSE versus speckle size for simulated fringe patterns with type $M_1$ fringe modulation }
	\label{fig:para_plot_speckle}
\end{figure}
\begin{figure}[h!]
	\centering
	\includegraphics[width=0.45\textwidth]{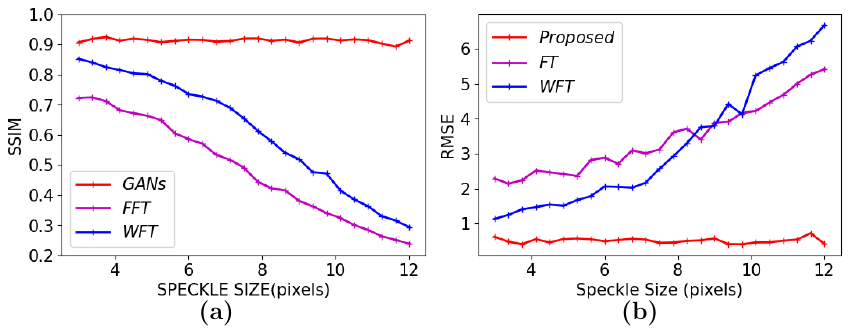} 
	\caption{(a) SSIM versus speckle size and (b) RMSE versus speckle size for simulated fringe patterns with type $M_2$ fringe modulation}
	\label{fig:sinu_plot_speckle}
\end{figure}

\section{Simulation Results}
\label{sec:sim}
For training purpose, we digitally simulated dataset consisting of 1,200 noisy fringe patterns and their corresponding normalized images.
Each fringe pattern has a size of 256 x 256 pixels. 
The phase map was modeled using Zernike polynomials with coefficients extending up to 20 orders, and spatial carrier frequency was incorporated with values ranging from 0.04 to 0.06 per pixel.
Additionally, the fringe background and modulation terms were modeled using Zernike polynomials, with dynamic range between 0 and 1.
Further speckle noise was introduced with speckle size varying between 4 and 8 pixels using the method described in \cite{kaufmann1996speckle}. 
In addition, we also added additive white Gaussian noise (AWGN) with a signal to noise ratio (SNR) ranging from 10 to 40 dB. 
To mitigate the risk of overfitting, we augmented the dataset through appropriate resizing and cropping operations \cite{isola2017image}. 
The optimizer was chosen to be Adam stochastic gradient descent algorithm \cite{kingma2014adam} with learning rate around 0.0002 and kept the value of multiplication factor $\lambda = 100$.
For training, we used a batch size of unity and number of epochs set to 15 in our model.
Python programming was used to perform the numerical simulations, and the neural network architecture was designed using Keras libraries \cite{geron2022hands}. 
Model training was done on workstation with 60 GB memory having NVIDIA RTX5000 graphic card.

To validate the effectiveness, the proposed method was tested for phase extraction from digitally simulated fringe patterns.
We also generated two types of non-linear fringe modulation maps for the fringe patterns, and designated them as types $M_1$ and $M_2$. 
We also added white Gaussian noise with an SNR of 0 dB to model severe noise conditions.
For a comparative quantitative assessment, phase estimation was also performed using state of the art techniques including the Fourier Transform (FT) method \cite{vinnichenko2023performance} and the Windowed Fourier Transform (WFT) method \cite{kemao2007two}.
For quantitative assessment, we also computed structural similarity index measure (SSIM) \cite{wang2004image, avanaki2009exact} and root mean square error (RMSE) for phase estimation using the given methods. 

The noisy fringe pattern is shown in Figure \ref{fig:para}(a).
The type $M_1$ fringe modulation map with sharp fluctuations is shown in Figure \ref{fig:para}(b). 
The original phase map (radians) is shown in Figure \ref{fig:para}(c).
The phase estimated using FT method is shown in Figure \ref{fig:para}(d) and its corresponding absolute phase estimation error map is shown in Figure \ref{fig:para}(e).
Similarly, phase calculated using WFT method is shown in Figure \ref{fig:para}(f) and corresponding estimation error is shown in Figure \ref{fig:para}(g).
The phase estimated using proposed method is shown in Figure \ref{fig:para}(h), and correspondingly, Figure \ref{fig:para}(i) displays the phase estimation error. 
For the FT method, the RMSE was about 2.2431 radians and SSIM metric was about 71.59\%.
Similarly, for the WFT method, the RMSE was about 1.0906 radians and SSIM metric was about 85.01\%.
Finally, for the proposed method, the RMSE was about 0.8206 radians and SSIM metric was about 89.30\%.

Furthermore, the type $M_2$ fringe modulation map is shown in Figure \ref{fig:sinu}(a).
The noisy fringe pattern is shown in Figure \ref{fig:sinu}(b). 
The phase map obtained from FT method is shown in Figure \ref{fig:sinu}(c), and accordingly,  Figure \ref{fig:sinu}(d) shows the related phase estimation error.
In addition, Figure \ref{fig:sinu}(e) depicts the phase estimated using WFT method, and Figure \ref{fig:sinu}(f) depicts the corresponding estimation error.
Finally, Figure \ref{fig:sinu}(g) displays the phase retrieved using proposed method, and Figure \ref{fig:sinu}(h) highlights the corresponding error.
For the FT method, the RMSE was about 2.0912 radians and SSIM metric was about 72.99\%.
Similarly, for the WFT method, the RMSE was about 1.0641 radians and SSIM metric was about 85.48\%.
Finally, for the proposed method, the RMSE was about 0.9042 radians and SSIM metric was about 88.66\%.

For further quantitative testing, we computed SSIM and RMSE values using these methods for wide range of noise levels.
For $M_1$ type fringe modulation, SSIM versus SNR plots using different methods are shown in Figure \ref{fig:para_plot}(a).
Further, Figure \ref{fig:para_plot}(b) shows RMSE versus SNR plots using the given methods. 
Similarly, for $M_2$ type fringe modulation, SSIM versus SNR plots using different methods are shown in Figure \ref{fig:sinu_plot}(a).
Further, Figure \ref{fig:sinu_plot}(b) shows RMSE versus SNR plots using the given methods. 

Next, to visualize the effectiveness of phase estimation methods in the presence of speckle noise, we digitally simulated fringe patterns with non-uniform fringe modulation and added white Gaussian noise (standard deviation 0.3645) as well as speckle noise (speckle size of 6 pixels) in our analysis. 
Figure \ref{fig:para_speckle}(a) shows the speckle noise corrupted fringe pattern corresponding to type $M_1$ fringe modulation.
The phase maps estimated using the FT method, WFT method and the proposed method are shown in parts (b,d,f) of Figure \ref{fig:para_speckle}.
Similarly, the phase estimation errors corresponding to the FT method, WFT method and the proposed method are shown in parts (c,e,g) of Figure \ref{fig:para_speckle}.
For the FT method, the RMSE was about 2.2481 radians and SSIM metric was about 69.51\%.
Similarly, for the WFT method, the RMSE was about 1.274 radians and SSIM metric was about 82.82\%.
Finally, for the proposed method, the RMSE was about 0.6401 radians and SSIM metric was about 90.85\%.

Similarly, Figure \ref{fig:sinu_speckle}(a) shows the speckle noise corrupted fringe pattern corresponding to type $M_2$ fringe modulation.
The phase maps estimated using the FT method, WFT method and the proposed method are shown in parts (b,d,f) of Figure \ref{fig:sinu_speckle}.
Similarly, the phase estimation errors corresponding to the FT method, WFT method and the proposed method are shown in parts (c,e,g) of Figure \ref{fig:sinu_speckle}.
In this case, for the FT method, the RMSE was about 2.5152 radians and SSIM metric was about 68.12\%.
Similarly, for the WFT method, the RMSE was about 1.4663 radians and SSIM metric was about 81.49\%.
Finally, for the proposed method, the RMSE was about 0.5495 radians and SSIM metric was about 91.23\%.

In addition, to better visualize the performance, the given methods were tested over different speckle sizes ranging from 3 to 12 pixels.
The SSIM and RMSE versus speckle size plots for type $M_1$ fringe modulation is shown in parts (a,b) of Figure \ref{fig:para_plot_speckle}.
Similarly, the SSIM and RMSE versus speckle size plots for type $M_2$ fringe modulation are shown in parts (a,b) of Figure \ref{fig:sinu_plot_speckle}.

\section{Experimental Results}
\label{sec:exp}
\begin{figure}[!t]
	\centering
	\includegraphics[width=0.49\textwidth]{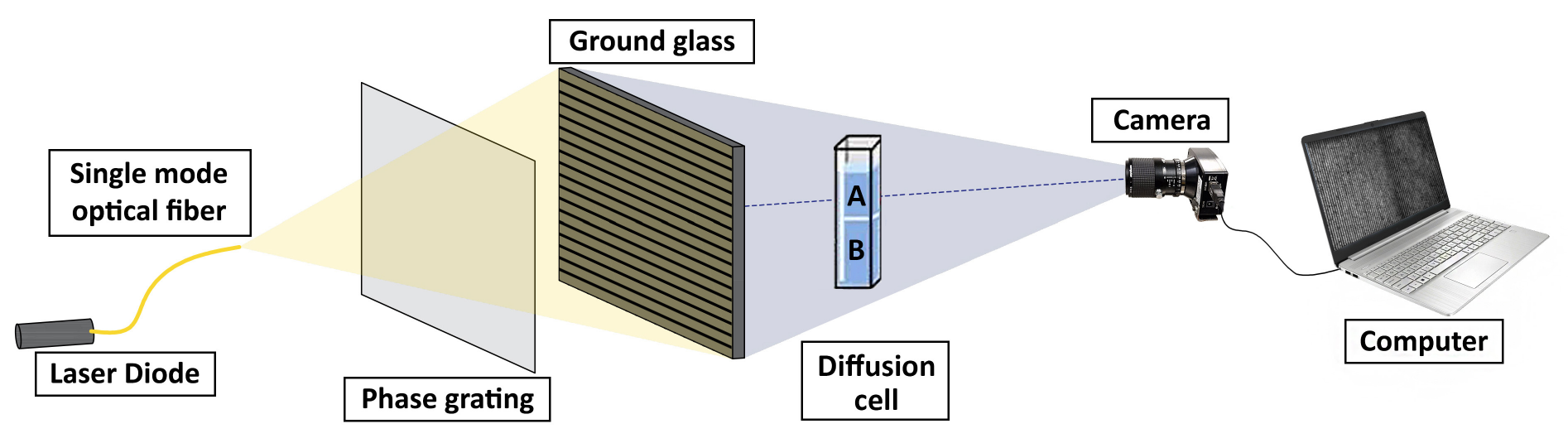} 
	\caption{Schematic of diffractive optical element BOS experimental setup }
	\label{fig:setup}
\end{figure}
\begin{figure}[t!]
    \centering
    {\includegraphics[width=0.49\textwidth]{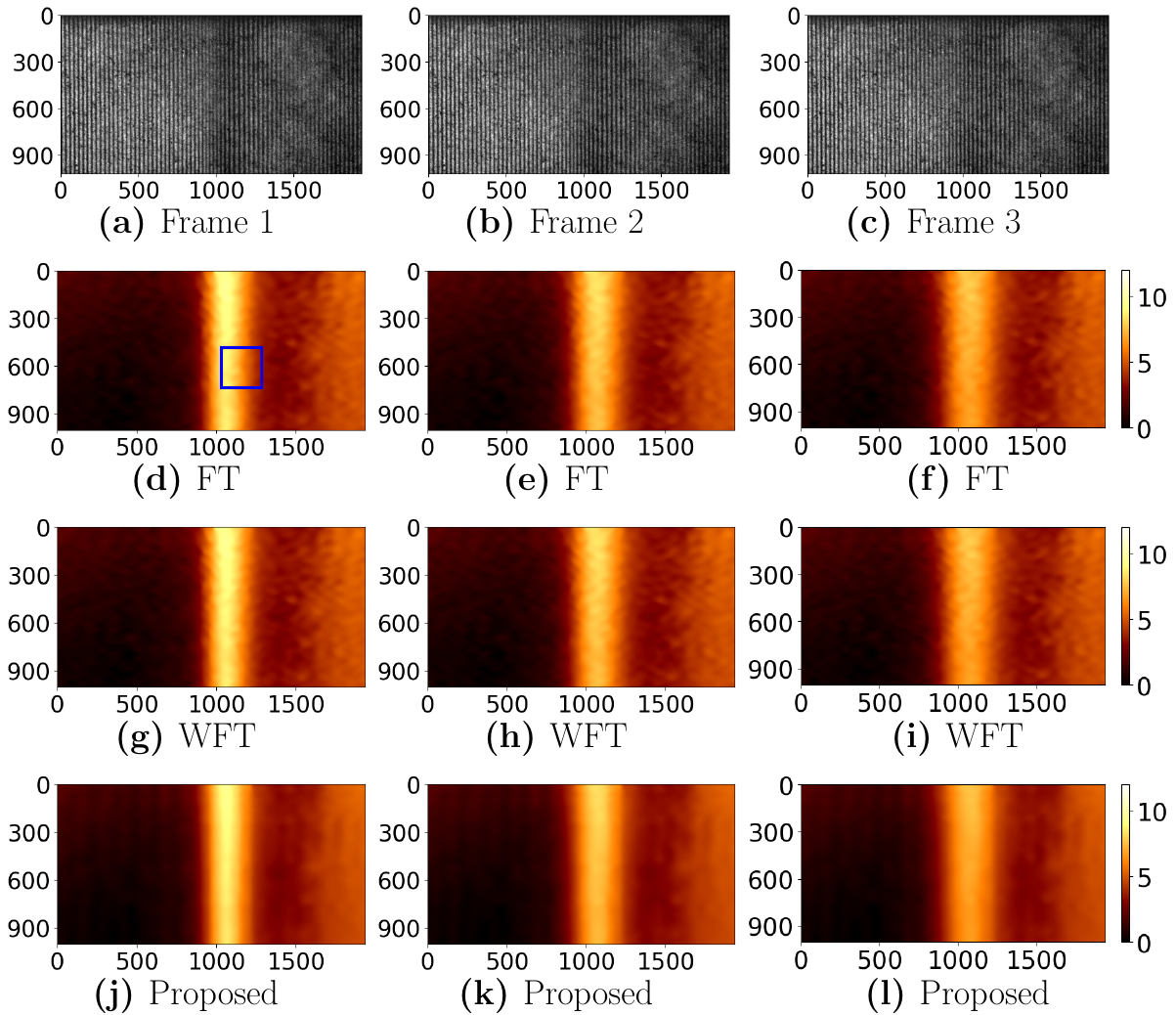}}
    \caption{
        (a-c) Experimental fringe patterns for frames 1-3.
    (d-f) Estimated phase using FT method.
    (g-i) Estimated phase using WFT method.
    (j-l) Estimated phase using proposed method.
    }
    \label{fig:expfri_1}
\end{figure}

\begin{figure}[t!]
    \centering
    {\includegraphics[width=0.49\textwidth]{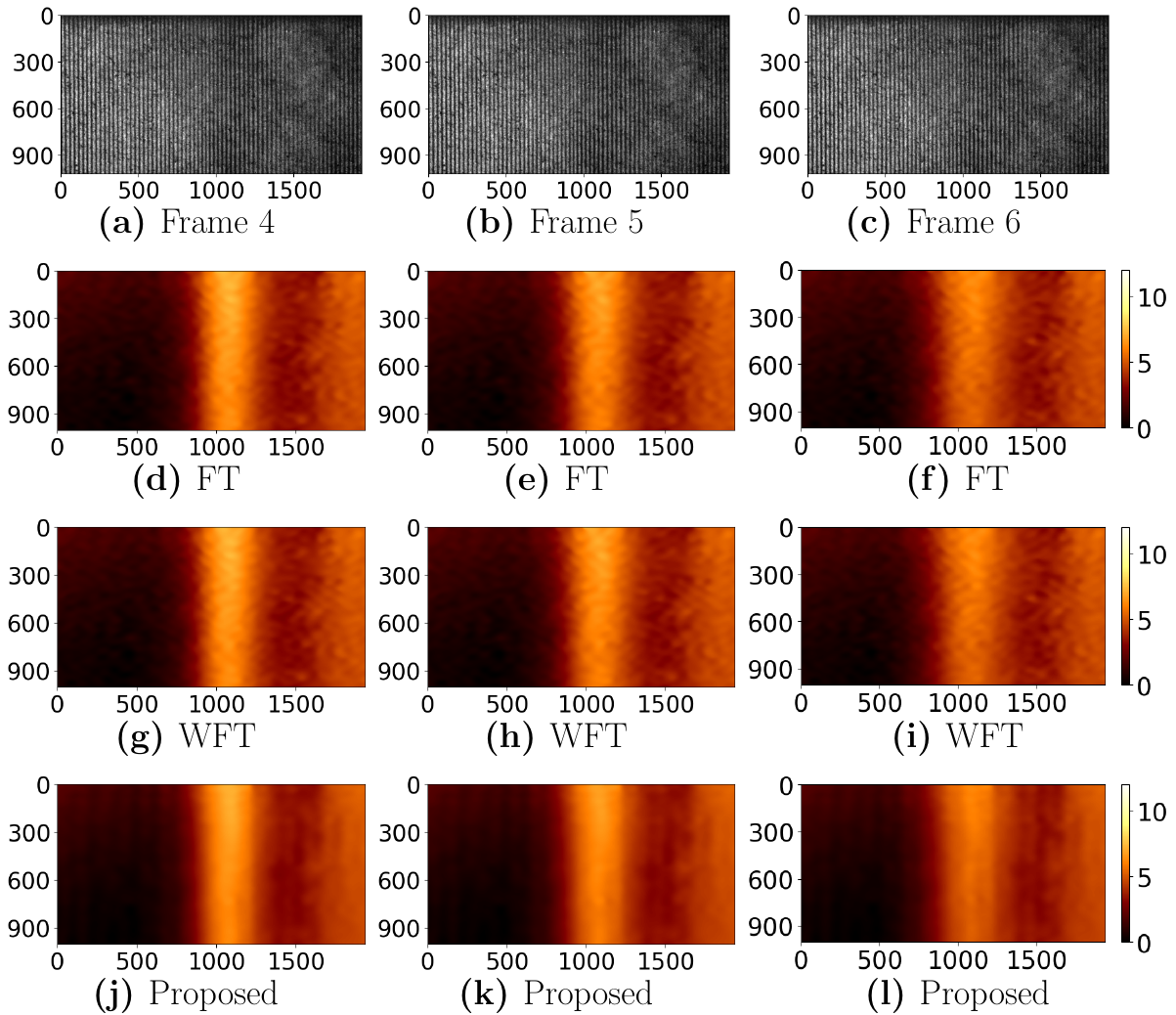}}
    \caption{
        (a-c) Experimental fringe patterns for frames 4-6.
    (d-f) Estimated phase using FT method.
    (g-i) Estimated phase using WFT method.
    (j-l) Estimated phase using proposed method.
}
    \label{fig:expfri_2}
\end{figure}
\begin{figure}[t!]
    \centering
    {\includegraphics[width=0.49\textwidth]{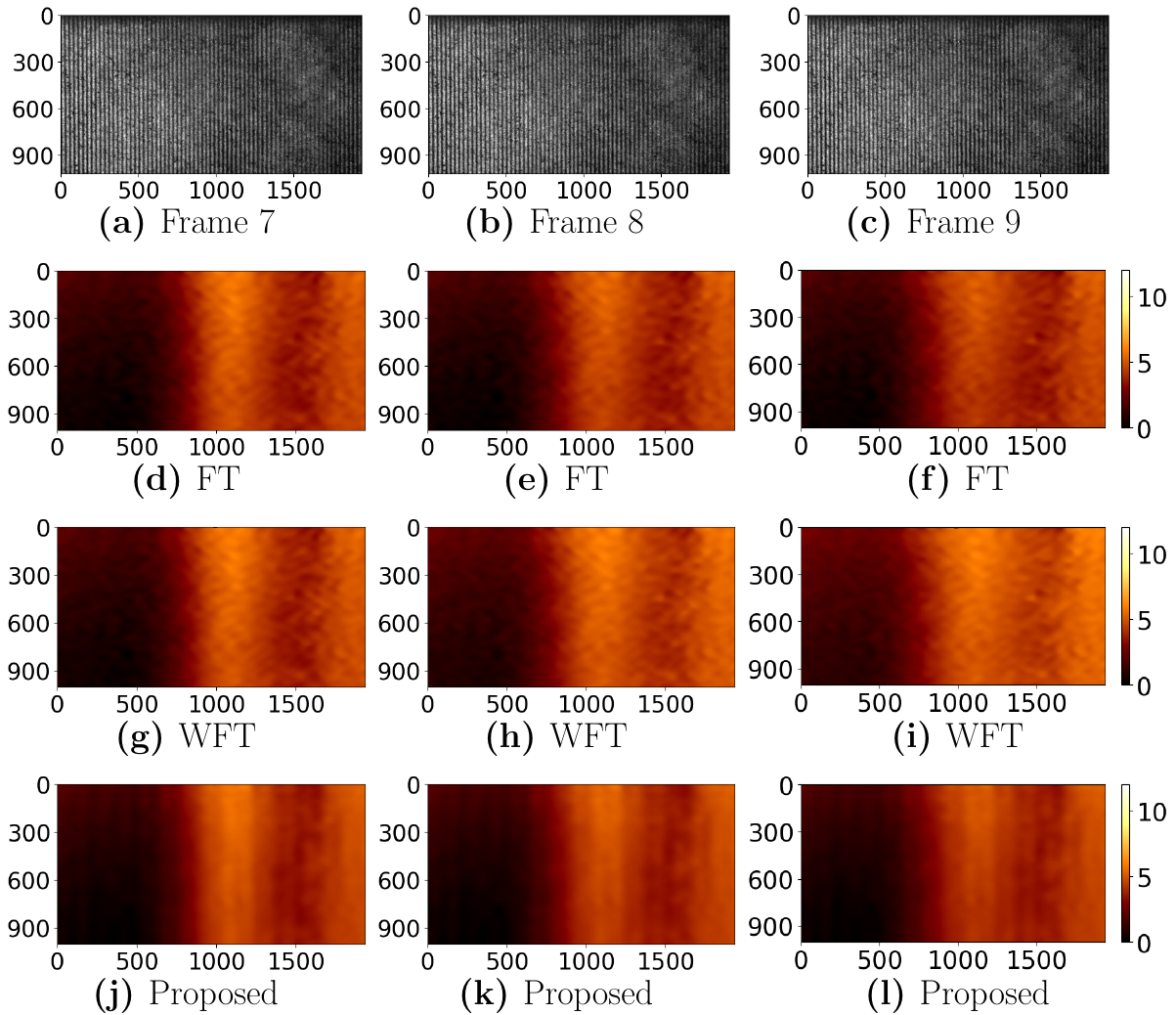}}
    \caption{
        (a-c) Experimental fringe patterns for frames 7-9.
    (d-f) Estimated phase using FT method.
    (g-i) Estimated phase using WFT method.
    (j-l) Estimated phase using proposed method.
}
    \label{fig:expfri_3}
\end{figure}
\begin{figure}[t!]
    \centering
    {\includegraphics[width=0.49\textwidth]{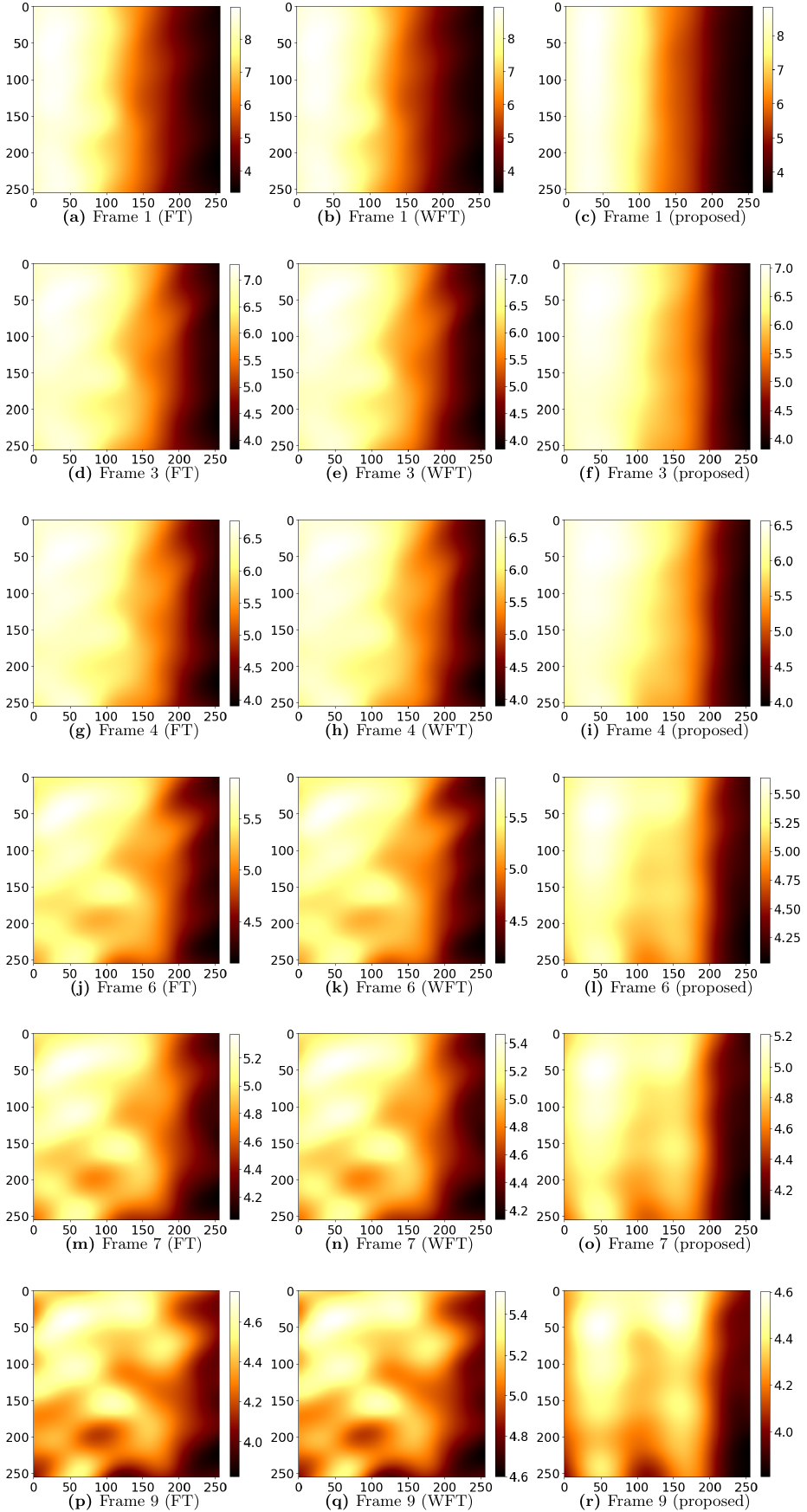}}
    \caption{
        (a,d,g,j,m,p) Zoomed phase maps using FT method for different frames.
        (b,e,h,k,n,q) Zoomed phase maps using WFT method for different frames.
        (c,f,i,l,o,r) Zoomed phase maps using proposed method for different frames.
}
    \label{fig:exp_mask}
\end{figure}

\begin{figure}[t!]
    \centering
    {\includegraphics[width=0.45\textwidth]{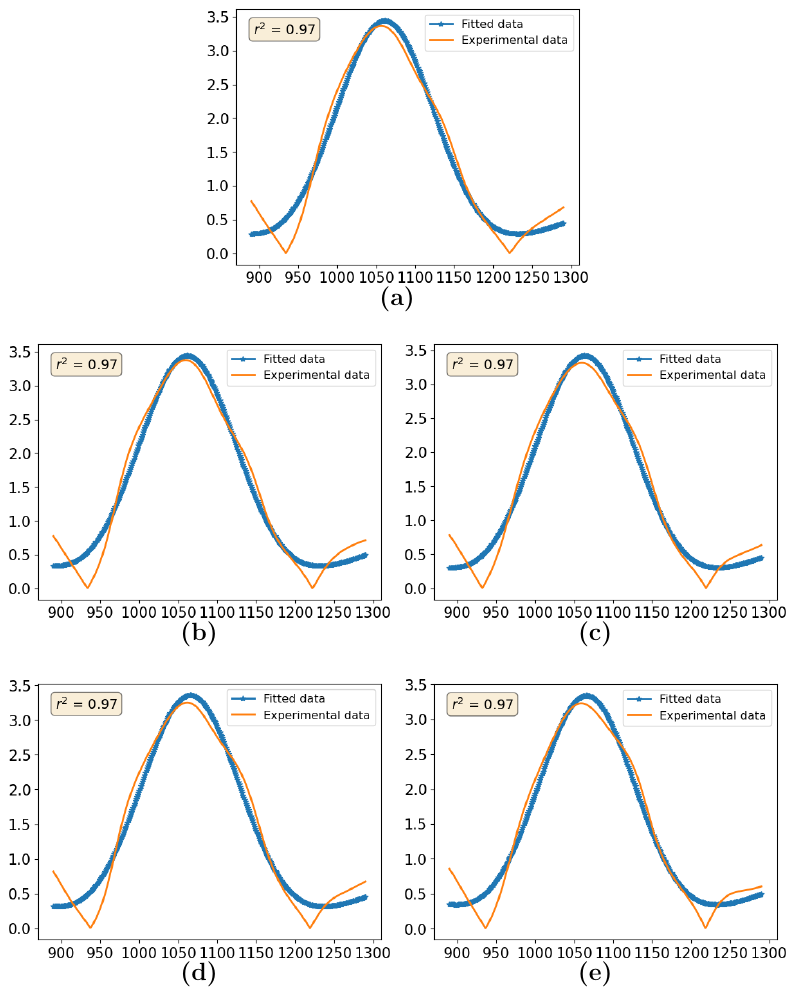}}
    \caption{
        Experimental $\Delta\phi$ and least squares fit around the diffusion region for row number (a) 150 (b) 350 (c) 502 (d) 650 and (e) 800.
}
    \label{fig:expfri_4}
\end{figure}

The schematic of the BOS experimental setup is shown in Figure \ref{fig:setup}.
For illumination, we used a fiber-coupled diode laser with wavelength of 670 nm in our setup.
The spherical wave emerging from the tip of the single mode optical fiber is incident on a sawtooth phase grating which serves as the diffractive optical element.
The diffraction efficiency of the grating is 0.31 for the +1 order and 0.4 for the zeroth order, implying that most of the incident power is distributed between these two orders. 
We also used a ground glass plate on which the grating pattern is projected, and its diffusive nature ensures good visibility of the projected pattern at the camera.
Our sample is a binary liquid solution comprising of pure water (A) and aqueous solution of common salt (B),  which is placed inside a spectrophotometric glass diffusion cell (10 mm $\times$ 8 mm) with a path length of 10 mm along the optical axis. 
In our setup, the imaging camera, a CMOS TV camera (Silicon Video 9T001C) equipped with TEC-55 55 mm F/2.8 Telecentric Computar Lens, captures the grating pattern or fringe pattern at various time intervals from the beginning of the diffusion process, thus resulting in a temporal stack of images. 
More details about the experimental setup are outlined in \cite{schirripa2004liquid}.
Due to the concentration gradient within the diffusion cell, the refractive index distribution remains non-uniform, which leads to spatially varying phase map in the recorded fringe pattern.

The effectiveness of the proposed method for practical applications is demonstrated through experimentally recorded fringe patterns in diffractive optical element-based background-oriented schlieren. 
The images (8-bit depth) were captured with an f-number of 4, an exposure time of 2.5 ms, and a magnification where 1 pixel corresponds to 9.1 $\mu m$ . 
The time-lapsed dataset consists of 9 sequentially acquired fringe patterns for the diffusion experiment, and we selected a common region of interest for analysis.
The first five frames were recorded at time intervals of 2 minutes whereas the next four frames were recorded at time intervals of 5 minutes.
From these experimental fringe patterns, we estimated the phase maps using the proposed method with batch-wise processing of neural network model.
For comparison, we also processed the fringe patterns using the FT method and WFT method.

The first three experimental fringe patterns are shown in Figure \ref{fig:expfri_1}(a,b,c).
The estimated phase maps using the FT method from these experimental fringe patterns are shown in parts (d-f) of Figure \ref{fig:expfri_1}.
Similarly, the estimated phase maps using the WFT method from the experimental fringe patterns are shown in parts (g-i) of Figure \ref{fig:expfri_1}.
Finally, the estimated phase maps using the proposed method from these experimental fringe patterns are shown in parts (j-l) of Figure \ref{fig:expfri_1}.
The next three experimental fringe patterns are shown in Figure \ref{fig:expfri_2}(a,b,c).
The estimated phase maps using the FT method from these experimental fringe patterns are shown in parts (d-f) of Figure \ref{fig:expfri_2}.
Similarly, the estimated phase maps using the WFT method from the experimental fringe patterns are shown in parts (g-i) of Figure \ref{fig:expfri_2}.
Finally, the estimated phase maps using the proposed method from these experimental fringe patterns are shown in parts (j-l) of Figure \ref{fig:expfri_2}.
Continuing, the final three experimental fringe patterns are shown in Figure \ref{fig:expfri_3}(a,b,c).
The estimated phase maps using the FT method from these experimental fringe patterns are shown in parts (d-f) of Figure \ref{fig:expfri_3}.
Similarly, the estimated phase maps using the WFT method from the experimental fringe patterns are shown in parts (g-i) of Figure \ref{fig:expfri_3}.
Finally, the estimated phase maps using the proposed method from these experimental fringe patterns are shown in parts (j-l) of Figure \ref{fig:expfri_3}.

For further clarity, we also show zoomed regions of the phase maps obtained using the different methods for frames 1,3,4,6,7 and 9 in Figure \ref{fig:exp_mask}.
The region of interest (marked by blue color) is shown in Figure \ref{fig:expfri_1}(d).
We used the same region to highlight the zoomed phase maps for the different methods across different frames.
In Figures \ref{fig:exp_mask}(a,d,g,j,m,p), we show the zoomed phase maps using FT method for different frames.
In Figures \ref{fig:exp_mask}(b,e,h,k,n,q), the zoomed phase maps using WFT method for different frames are shown.
Finally, Figures \ref{fig:exp_mask}(c,f,i,l,o,r) display zoomed phase maps using proposed method for different frames.
From the figures, the variations in phase maps across different frames is evident.

The experimental results clearly illustrate the temporal variation in phase distribution which is caused by the refractive index variations introduced by the diffusion process.
Notably, with the progression of diffusion phenomenon over time, the non-uniformity of the refractive index for the initial mixture stabilizes towards a uniform distribution, which is also evident from the shown phase maps.
Next, it has been shown that the diffusion coefficient for the temporal flow process can be mapped to the temporal phase difference between the frames \cite{schirripa2004liquid}.
Further, as the diffusion occurs mainly along horizontal dimension $x$, the phase difference was mainly computed along the image columns for five different rows with 150, 300, 502 (middle row), 650 and 800 as the respective row numbers.
Equivalently, we computed the temporal phase difference map $\Delta\phi(x,t_1,t_2) = \phi(x,y,t_1) - \phi(x,y,t_2)$ by considering the first frame ($t_1=120$ seconds) and sixth frame ($t_2=900$ seconds).
It has been shown that the temporal phase difference map and Diffusion coefficient ($D$) are mathematically related as \cite{schirripa2004liquid}
\begin{equation}
    \Delta\phi(x,t_1,t_2) \propto \frac{\exp[-x^2/4Dt_1]}{2\sqrt{Dt_1}} -  \frac{\exp[-x^2/4Dt_2]}{2\sqrt{Dt_2}}   
\end{equation}
where the initial point of separation of two liquids is marked as $x=0$.
Accordingly, we used the above equation for computing the least squares fit for the experimental phase difference map near the diffusion region (around 900-1300 pixels) for a given row, and repeated the procedure for the five different rows.
These results are shown in Figure \ref{fig:expfri_4} along with the coefficient of determination parameter $r^2$, which acts as a measure of goodness of fit \cite{chicco2021coefficient}.
Using the curve-fitting procedure, the diffusion coefficient was estimated for the five rows, and we computed the mean and standard deviation for these values.
Accordingly, we obtained the diffusion coefficient estimate as $D = 1.429 \times 10^{-9}$ (mean) $\pm$ $0.024\times 10^{-9}$ (standard deviation) m$^2$/s in our analysis. 
This value agrees well with the reported value  of $1.47 \times 10^{-9}$ m$^2$/s for the diffusion coefficient \cite{ambrosini2014diffusion}.
These results show the potential of deep learning for quantitative visualization of diffusive flows in BOS setup.

\section{Conclusions}

We proposed a deep learning assisted signal subspace method for reliable phase estimation from fringe patterns obtained in diffractive optical element based BOS setup.
The simulation results clearly demonstrate that the proposed method effectively estimates phase even under extreme noise and nonlinear fringe modulation fluctuations. 
Furthermore, quantitative assessment metrics based on SSIM and RMSE highlight its superior performance when compared to state of the art methods. 
Notably, even for severe noise levels, with SNR values in the range 0-5 dB, the RMSE values provided by the proposed method lie within 1 radian, as shown in Figures \ref{fig:para_plot} and \ref{fig:sinu_plot}.
Similarly, RMSE values obtained using proposed method are less than 1 radian even with varying speckle sizes, as is evident from Figures \ref{fig:para_plot_speckle} and \ref{fig:sinu_plot_speckle}. 
In addition, the SSIM values obtained using proposed method are consistently greater than 90\% for severe noise levels (as shown in Figures \ref{fig:para_plot} and \ref{fig:sinu_plot}) and varying speckle sizes (as shown in Figures \ref{fig:para_plot_speckle} and \ref{fig:sinu_plot_speckle}) in our analysis.
These results indicate high tolerance of the proposed method against noise.
Experimental results further validate the practical applicability of the proposed method.
Overall, we believe that our method offers a step towards inclusion of deep learning methodology for flow visualization.


\end{document}